\documentclass[english,twocolumn, fleqn]{article}
\usepackage[T1]{fontenc}
\usepackage[latin9]{inputenc}
\usepackage[letterpaper]{geometry}
\usepackage{color}
\usepackage{amsmath}
\usepackage{graphicx}
\usepackage{amssymb}
\usepackage{esint}
\usepackage[numbers]{natbib}

\makeatletter
\newcommand{\lyxaddress}[1]{
\par {\raggedright #1
\vspace{1.4em}
\noindent\par}
}

\makeatother

\usepackage{babel}

\begin{document}

\title{Two-cluster bifurcations in systems of globally pulse-coupled oscillators}

\author{Leonhard Lücken, Serhiy Yanchuk}

\maketitle

\lyxaddress{Institute of Mathematics, Humboldt University of Berlin, Unter den
Linden 6, 10099 Berlin, Germany}
\begin{abstract}
For a system of globally pulse-coupled phase-oscillators, we derive
conditions for stability of the completely synchronous state and all
possible two-cluster states and explain how the different states are
naturally connected via bifurcations. The coupling is modeled using
the phase-response-curve (PRC), which measures the sensitivity of
each oscillator's phase to perturbations. For large systems with a
PRC, which turns to zero at the spiking threshold, we are able to
find the parameter regions where multiple stable two-cluster states
coexist and illustrate this by an example. In addition, we explain
how a locally unstable one-cluster state may form an attractor together
will its homoclinic connections. This leads to the phenomenon of intermittent,
asymptotic synchronization with abating beats away from the perfect
synchrony.
\end{abstract}

\section{Introduction}

Ensembles of interacting dynamical systems appear as mathematical
models in many branches of science \citep{Pikovsky2001,Strogatz2001}.
For example, the study of coupled lasers \citep{Wunsche2005,Fischer2006,Yanchuk2004,Yanchuk2006}
is triggered by technological applications such as high-power generation
or secure communication \citep{Cuomo1993,Kanter2008}. Interacting
biological units play an important role for the functioning of living
organisms \citep{Mosekilde2002}. Mechanical or electrical coupled
oscillators have been extensively studied as paradigmatic models \citep{Perlikowski2008,Perlikowski2010}
to study various synchronization phenomena. In neuroscience, the synchronization
of neuron populations plays an important role \citep{Tass1999,Timme2006,Popovych2006}
and might even lead to pathological effects \citep{Elble1990}. As
a result, there was recently an increasing effort to control the desynchronization
of populations of coupled oscillators. In particular, the coordinated
reset stimulation technique \citep{Tass1999,Tass2003} proposes to
establish a cluster-state in the network, in which the oscillator's
phases split into several subgroups. This example illustrates the
importance of the analysis of cluster formation in coupled systems. 

In this paper we investigate the connection between the properties
of single oscillators and their influence on the appearance of clusters
in a globally coupled system of such oscillators. We consider the
pulse coupling \citep{Goel2002,Mirollo1990,Bottani1996,LaMar2010,Bressloff1997}
approximation, which is widely used for modeling of neuron populations.
Such an approximation is appropriate in the case, when the interaction
time between the oscillators is much smaller than the characteristic
period of oscillations. The interaction is mediated by the pulses,
which are emitted by each of the oscillator after reaching some threshold.
Although the coupling topology is fixed and assumed to be global,
i.e. all-to-all and homogeneous, the dynamical properties of the individual
oscillators will be considered variable (all at the same time). This
enables us to describe a bifurcation scenario, which links dynamic
regimes of stable synchrony with regimes, that promote cluster-formation.
More specifically, we consider various sensitivity functions for the
individual oscillators, called the phase-response-curve \citep{Goel2002,Brown2004}.
PRCs have been introduced, studied, and computed for many neuronal
models \citep{Winfree2001,Ermentrout1996,Hoppensteadt1997,Brown2004,Novicenko2011}.
They can serve as an appropriate control parameter determining the
properties of the oscillators in the network\textcolor{black}{{} \citep{Achuthan2009}.
In our paper, we restrict our analysis to PRCs, }which turn to zero
at the threshold, at which the oscillator emits a pulse. This choice
is motivated by several well known neuron models \citep{Goel2002,Brown2004,Izhikevich2005}
and means that the uncoupled system at the threshold is insensitive
to small external perturbations.

Our work is accomplished by the bifurcation analysis of the cluster
states and the analysis of the spectral properties of the completely
synchronous state and eventually emerging two-cluster states. Our
results reveals how the synchronization properties of the network
depend on the PRC of the individual oscillators. We point out how
the loss of stability of the synchronous solution may give rise to
stable homoclinic connections of the synchronous solution and invariant
two clusters. We derive conditions for the stability of these states
and, in an exemplary family of unimodal, positive PRCs, we illustrate
the mechanism by shifting the position of their maxima. For some range
of the control parameter we observe a stable homoclinic connection
of the synchronous solution, which leads to an apparent synchronization
of the population with eventual beats out of perfect sync at a large
time-scale, which are becoming more rare with time. Thus, practically,
a synchronized state is observed generically in such systems even
in the case, when the completely synchronous state is locally unstable.
For another range of parameters, we observe the stabilization of various
two-cluster states, which bifurcate from the completely synchronous
one-cluster state as predicted by the analytics. In the paper, the
notions ''complete synchronization'' and ''one-cluster'' solution
will be used interchangeably as they have the same meaning for our
model.

The structure of the paper is as follows. In Sec.~\ref{sub:The-system},
the system is introduced. In Sec.~\ref{sec:Invariant-cluster-subspaces},
we reduce the dynamics to one-dimensional maps and in Sec.~\ref{sec:one-cluster},
we review the stability and bifurcations of the one-cluster state
in this framework. We also explain how the existence of a stable homoclinic
set may lead to the phenomenon of intermittent asymptotic synchronization.
Appearance and stability of two-cluster states are studied in Sec.~\ref{sub:bifurcations-of-the-synchronous-state}
and \ref{sec:two-cluster}. Numerics for some illustrative example
is shown in Sec.~\ref{sec:Numerics} and some technical details are
included in the Appendix.

\section{\label{sub:The-system} The system}

Networks of weakly coupled oscillators can often be reduced to phase-models,
which keep a single phase variable $\varphi\in\left[0,2\pi\right]$
for each oscillator \citep{Kuramoto1984,Brown2004,Hoppensteadt1997}.
In this reduction, one naturally encounters a function which measures
the local sensitivity of an oscillator's phase to small perturbations
-- the phase response curve (PRC). If the perturbations act along
only one direction of the state space, e. g. the voltage component
in many neuron models, the PRC can be represented as a scalar function.
Further reduction is possible when the coupling takes place on a considerably
smaller time scale than the period of oscillations. In this case it
is reasonable to approximate the interaction by an impact, i.e. by
assuming that the interaction is immediate. By combining these two
reductions, one obtains the model of pulse-coupled phase-oscillators
\citep{Strogatz1992,Goel2002,Memmesheimer2006,Achuthan2009}, which
is the subject of our paper.

\begin{figure}
\includegraphics[width=0.45\textwidth]{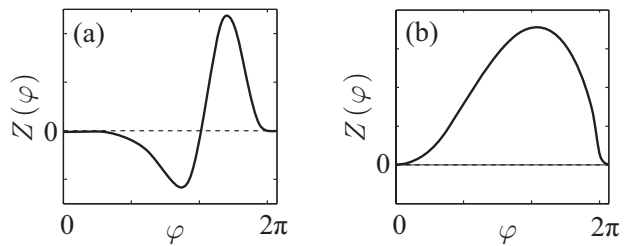}\caption{\label{fig:PRC-examples}Examples of different PRCs. (a) Hodgkin-Huxley
model, (b) Connor model. (adapted from \citep{Ermentrout1996}).}

\end{figure}

We consider the system of $N$ identical phase-oscillators, whose
dynamics are given as\begin{align}
\dot{\varphi}_{j}\left(t\right) & =1+\frac{\varkappa}{N}Z\left(\varphi_{j}\left(t\right)\right)\sum_{t_{kl},k\ne j}\delta(t-t_{kl}),\ j=1,...,N.\label{eq:system-def-1}\end{align}
Here $t_{kl}$ are time moments when $k$-th oscillator reaches the
threshold $\varphi_{k}(t_{kl})=2\pi$, $l=1,2,\dots$. We call these
time moments also ''spikes''. At this time, the $k$-th oscillator
sends a spike and all other oscillators $j$ with $j\ne k$ obtain
an impact\begin{align}
\varphi_{j}\mapsto\mu\left(\varphi_{j}\right) & :=\varphi_{j}+\frac{\varkappa}{N}Z\left(\varphi_{j}\right),\label{eq:mu}\end{align}
where $Z(\varphi)$ is the PRC. At the same time moment, the $k$-th
oscillator resets to $\varphi_{k}=0$. It is assumed that the phase
response curve $Z\left(\varphi\right)$ is smooth inside $\left(0,2\pi\right),$
but smoothness of derivatives in the endpoints is explicitly not assumed,
i.e. possibly $Z^{\prime}\left(0\right)\neq Z^{\prime}\left(2\pi\right).$
Moreover, we assume $Z\left(0\right)=Z\left(2\pi\right)=0,$ which
is frequently met in neuron models and means that a neuron is insensitive
to external stimulation when it is generating or has just generated
an action potential \citep{Goel2002,Brown2004,Izhikevich2005} (see
Fig.~\ref{fig:PRC-examples}). $\varkappa>0$ is the coupling strength.

An obvious but important property of system (\ref{eq:system-def-1})
is that the order of the phases is invariant if the quotient $\varkappa/N$
of coupling strength and network size is sufficiently small. This
means that the oscillators do not overrun each other with time. If
\begin{align}
\frac{\varkappa}{N} & <\frac{1}{\left|\min_{\varphi\in\left[0,2\pi\right]}Z'\left(\varphi\right)\right|},\label{eq:phaseordercond}\end{align}
the map (\ref{eq:mu}) becomes strictly monotonous and thus, the phase-ordering
is preserved during spikes as well.

Although the dynamical system (\ref{eq:system-def-1}) together with
the resetting condition is defined completely, it is worth to notice
that it is equivalent to a discrete $N$-dimensional return map $(\varphi_{1},\dots,\varphi_{N})\to K(\varphi_{1},\dots,\varphi_{N})$.
This return map can be obtained by fixing the position of one oscillator,
e.g. $\varphi_{N}=0$. Taking into account that the phase ordering
is preserved, one iteration of the return map corresponds to one spike
of each oscillator, i.e. $N$ spikes altogether. More exact definition
of the return map is given in Appendix~\ref{sub:return-map}.

Given (\ref{eq:phaseordercond}), the synchronous solution\begin{align*}
\varphi_{1} & =\cdots=\varphi_{N}\end{align*}
is known to be linearly stable if and only if for all $N_{1}<N$ the
following condition holds \begin{align}
\left|\left(1+\frac{\varkappa}{N}Z^{\prime}\left(0\right)\right)^{N-N_{1}}\left(1+\frac{\varkappa}{N}Z^{\prime}\left(2\pi\right)\right)^{N_{1}}\right| & <1.\label{eq:linstab1cl}\end{align}
This was proven in \citep{Goel2002} by analyzing the linearization
of the return-map. The terms in (\ref{eq:linstab1cl}) are the eigenvalues
of this linearization at its fixed point $\varphi_{1}=...=\varphi_{N}.$
Remarkably, for the case $Z^{\prime}\left(0\right)\neq Z^{\prime}\left(2\pi\right),$
there might be a loss of stability induced by an increasing of the
population size.

\section{Dynamics in invariant cluster subspaces\label{sec:Invariant-cluster-subspaces}}

\begin{figure}
\includegraphics[width=0.45\textwidth]{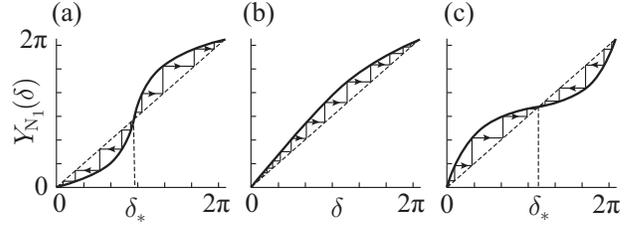}\caption{\label{fig:fixpt-and-cobweb-for-Y}Cobweb-diagrams for the one-dimensional
maps $Y_{N_{1}}$ within the two-cluster subspaces $\Pi_{N_{1}}$.
A fixed point $\delta_{\ast}\in\left(0,2\pi\right)$ corresponds to
a stationary two-cluster state of the return map. Chart (a) shows
a stable synchronous state and an unstable two-cluster state. In (b)
one can observe a homoclinic connection of the synchronous state,
resembled by a heteroclinic connection of $\delta=0$ and $\delta=2\pi$
under $Y_{N_{1}},$ and (c) shows an unstable synchronous state together
with a two-cluster state, that is stable to perturbations in the subspace
$\Pi_{N_{1}}$. Possible situations with several fixed points for
$Y_{N_{1}}$ are not shown. In Sec.~\ref{sub:bifurcations-of-the-synchronous-state},
we explain, how the cases (a) and (b), resp. (b) and (c), are connected
via the bifurcations of the synchronous solution.}

\end{figure}

In this section, we study the dynamics of the return map in invariant
subspaces $\Pi_{N_{1}}$ of the form \begin{align}
\Pi_{N_{1}} & =\left.\left\{ \begin{array}{r}
\varphi_{1}=\dots=\varphi_{N_{1}}=\delta,\\
\varphi_{N_{1}+1}=\dots=\varphi_{N}=0\end{array}\right|\delta\in\left(0,2\pi\right)\right\} .\label{eq:two-cluster}\end{align}
In these subspaces, the population is split into two clusters. The
distance between those clusters is denoted by $\delta$ (or $2\pi-\delta$),
which is the natural coordinate in $\Pi_{N_{1}}.$ We can reduce the
action of return map on $\Pi_{N_{1}}$ as follows. Choose a particular
initial state in $\Pi_{N_{1}},$ i.e.\[
\varphi_{1}=\dots=\varphi_{N_{1}}=\delta\mbox{ and }\varphi_{N_{1}+1}=\dots=\varphi_{N}=0,\]
with a particular initial distance $\delta.$ Under the dynamics of
(\ref{eq:system-def-1}), the next event in time will be the collective
burst of the oscillators of the first cluster at time $t=2\pi-\delta$.
These are immediately resetted to $\varphi=0,$ i. e.\[
\varphi_{1}=...=\varphi_{N_{1}}=0.\]
The impact of the burst on an oscillator of the second cluster is
obtained by repetitive application of the function $\mu$ from (\ref{eq:mu})
-- one time for each oscillator in the first cluster. This leads to
\[
\varphi_{N_{1}+1}=...=\varphi_{N}=\mu^{N_{1}}\left(2\pi-\delta\right),\]
where $\mu^{n}$ denotes the $n$-fold superposition of the function
$\mu$, i.e. \linebreak{}
 $\mu^{n}(\delta)=\underbrace{\mu(\mu(\mu(\cdots\mu}_{n}(\delta)\cdots))).$
At next, all oscillators advance with equal velocity, until the second
cluster reaches the threshold, that is,\[
\varphi_{N_{1}+1}=...=\varphi_{N}=2\pi.\]
Accordingly the first cluster is located at \[
\varphi_{1}=...=\varphi_{N_{1}}=2\pi-\mu^{N_{1}}\left(2\pi-\delta\right).\]
The following burst of the second cluster completes the reduction
of the return map and after the application of the return map, the
new distance is given by\begin{align}
Y_{N_{1}}\left(\delta\right) & =\mu^{N-N_{1}}\left(2\pi-\mu^{N_{1}}\left(2\pi-\delta\right)\right),\quad\delta\in\left[0,2\pi\right].\label{eq:def-YN1}\end{align}
The one-dimensional map (\ref{eq:def-YN1}) determines completely
the dynamics within the subspace $\Pi_{N_{1}}$, i.e. the dynamics
of perfect two-clusters (\ref{eq:two-cluster}). Fig.~\ref{fig:fixpt-and-cobweb-for-Y}
shows typical behaviors of these functions and corresponding cobweb-diagrams.
A fixed point $\delta_{*}\in\left(0,2\pi\right)$ of this map correspond
to a two-cluster fixed point of the global return map $K$ or, equivalently,
to a periodic two-cluster state of the original system (\ref{eq:system-def-1}).
In general, the shape of $Y_{N_{1}}(\delta)$ depends on the cluster
sizes $(N_{1},N-N_{1})$, the PRC $Z\left(\varphi\right)$, and the
coupling strength $\varkappa$.

In particular, these subspaces are of interest, since bifurcations
along them govern the stability properties of the synchronous state,
as we show in the next section.

\section{\label{sec:one-cluster} Stability of the synchronous state}

The trivial fixed points $\delta=0$ and $\delta=2\pi$ of $Y_{N_{1}}$
correspond to the synchronous state. Derivatives of $Y_{N_{1}}$ at
these points can be computed as\begin{align}
Y_{N_{1}}^{\prime}\left(0\right) & =\left(1+\frac{\varkappa}{N}Z^{\prime}\left(0\right)\right)^{N-N_{1}}\left(1+\frac{\varkappa}{N}Z^{\prime}\left(2\pi\right)\right)^{N_{1}},\nonumber \\
Y_{N_{1}}^{\prime}\left(2\pi\right) & =\left(1+\frac{\varkappa}{N}Z^{\prime}\left(0\right)\right)^{N_{1}}\left(1+\frac{\varkappa}{N}Z^{\prime}\left(2\pi\right)\right)^{N-N_{1}}\nonumber \\
 & =Y_{N-N_{1}}^{\prime}\left(0\right).\label{eq:DYN1-in-0}\end{align}
Since the linearized dynamics of the cluster distance $\delta$ for
small $\delta>0$ is governed by \begin{equation}
\delta\to Y_{N_{1}}^{\prime}\left(0\right)\delta,\label{eq:linz}\end{equation}
the derivatives $Y_{N_{1}}^{\prime}\left(0\right)$ are multipliers
of the map governing its local stability at $\delta=0$. Hence the
linear stability conditions of the synchronous state $\delta=0$ under
$Y_{n}$ for $n=1,...,N-1$ are $|Y_{n}'(0)|<1.$ These are identical
to (\ref{eq:linstab1cl}) because the invariant directions $\varphi_{1}=...=\varphi_{N_{1}}=1,$
$\varphi_{N_{1}+1}=...=\varphi_{N}=0$ are eigenvectors of the return
map at the synchronous solution $\varphi_{1}=...=\varphi_{N}=0$.
It is evident, that the maximal multiplier of the linearized map (\ref{eq:linz})
is either \[
Y_{1}^{\prime}\left(0\right)=\left(1+\frac{\varkappa}{N}Z^{\prime}\left(0\right)\right)^{N-1}\left(1+\frac{\varkappa}{N}Z^{\prime}\left(2\pi\right)\right)\]
 or \[
Y_{N-1}^{\prime}\left(0\right)=\left(1+\frac{\varkappa}{N}Z^{\prime}\left(0\right)\right)^{1}\left(1+\frac{\varkappa}{N}Z^{\prime}\left(2\pi\right)\right)^{N-1}\]
depending on the values of $Z'(0)$ and $Z'(2\pi)$. This means, that
the maximal growth of the cluster distance is observed in the case
when only one oscillator is separated from the rest of the population,
i.e. $N_{1}=1$ or $N_{1}=N-1$. 

When the number of oscillators $N$ is large, one can obtain the approximation\begin{align}
Y_{N_{1}}^{\prime}\left(0\right) & \approx\exp\left[\varkappa\left(\left(1-p\right)Z^{\prime}\left(0\right)+pZ^{\prime}\left(2\pi\right)\right)\right],\label{eq:limitN-of-YN1}\end{align}
where $p=N_{1}/N$ denotes the ratio of the oscillator number in the
first cluster to the total number. This means that for large populations,
the stability of the synchronous state to perturbations in form of
two-clusters with $N_{1}=pN$ and $N-N_{1}=\left(1-p\right)N$ elements
is governed by the condition\begin{align}
\left(1-p\right)Z^{\prime}\left(0\right)+pZ^{\prime}\left(2\pi\right) & <0\label{eq:split-stability}\end{align}
and is independent on coupling strength $\varkappa$. In particular,
the synchronous solution is linearly stable for large populations
only if both $Z^{\prime}\left(2\pi\right)<0$ and $Z^{\prime}\left(0\right)<0.$

\section{Two-cluster bifurcations of the synchronous state\label{sub:bifurcations-of-the-synchronous-state}}

In this section, we study generic codimension-1 bifurcations of the
synchronous state, which lead to the emergence of various two-cluster
states. We assume that the PRC is varied smoothly in some parameter
$\beta,$ that is, \[
Z\left(\varphi\right)=Z_{\beta}\left(\varphi\right).\]
In turn, the functions $\mu\left(\varphi\right)=\mu\left(\varphi,\beta\right)$
and $Y_{N_{1}}\left(\delta\right)=Y_{N_{1}}\left(\delta,\beta\right)$
become $\beta$-dependent, too. For the sake of readability, we will
mostly omit this dependency in formulas. We still demand that $Z_{\beta}\left(0\right)=Z_{\beta}\left(2\pi\right)=0$
and that (\ref{eq:phaseordercond}) is fulfilled with $Z=Z_{\beta}$
for all $\beta.$ 

As seen in the previous section, the synchronous solution is linearly
stable in the two-cluster subspace $\Pi_{N_{1}}$ if and only if

\begin{align}
\max\left\{ \lambda_{N_{1}}\left(\beta\right),\lambda_{N-N_{1}}\left(\beta\right)\right\}  & <1,\label{eq:inter-cluster-stab-01}\end{align}
where \begin{align*}
\lambda_{n}\left(\beta\right) & =Y_{n}^{\prime}\left(0\right)=Y_{N-n}'(2\pi)\\
 & =\left(1+\frac{\varkappa}{N}Z_{\beta}^{\prime}\left(0\right)\right)^{n}\left(1+\frac{\varkappa}{N}Z_{\beta}^{\prime}\left(2\pi\right)\right)^{N-n}.\end{align*}
The loss of stability of the synchronous state in the subspace $\Pi_{N_{1}}$
takes place if one of the multipliers $\lambda_{N_{1}}\left(\beta\right)$
or $\lambda_{N-N_{1}}\left(\beta\right)$ becomes bigger than $1$
as $\beta$ passes some critical value $\beta_{1}$ with \[
\max\left\{ \lambda_{N_{1}}\left(\beta_{1}\right),\lambda_{N-N_{1}}\left(\beta_{1}\right)\right\} =1.\]
Such bifurcation causes the change of the form of the mapping $Y_{N_{1}}$
from those shown in Fig.~\ref{fig:fixpt-and-cobweb-for-Y}(a) to
Fig.~\ref{fig:fixpt-and-cobweb-for-Y}(b). Note that, generally,
this loss is only one sided, giving the possibility of an emergence
of a robust homoclinic connection to the synchronous solution as seen
in Fig.~\ref{fig:fixpt-and-cobweb-for-Y}(b). In a second bifurcation,
the stability might be lost completely as $\beta$ reaches a critical
value $\beta_{2}$ with \[
\min\left\{ \lambda_{N_{1}}\left(\beta_{2}\right),\lambda_{N-N_{1}}\left(\beta_{2}\right)\right\} =1.\]
If $\beta_{1}\neq\beta_{2},$ there is either a creation or annihilation
of a fixed point $\delta_{\ast}=Y_{N_{1}}\left(\delta_{\ast}\right)\in\left(0,2\pi\right)$
in each critical value of $\beta$. Note that, since $Y_{n}^{\prime}\left(2\pi\right)=Y_{N-n}^{\prime}\left(0\right),$
the bifurcation points $\beta=\beta_{c}$ will coincide for $N_{1}=n$
and $N_{1}=N-n.$ The existence of a fixed point $\delta_{1}$ of
$Y_{N_{1}}$ always implies that $\delta_{2}=\mu^{N_{1}}\left(2\pi-\delta_{1}\right)$
is a fixed point of $Y_{N-N_{1}}.$ Therefore, the bifurcations appear
as pitchfork bifurcations of the return map.

\section{Stability of two-cluster states \label{sec:two-cluster}\label{sub:2Cl-stab-formula}}

In this section, we will provide explicit conditions for the stability
of the two-cluster state 

\begin{equation}
\varphi_{1}=...=\varphi_{N_{1}}=\delta_{\ast},\ \varphi_{N_{1}+1}=...=\varphi_{N}=0.\label{eq:2Cl-state}\end{equation}
These conditions will be given by the formulas (\ref{eq:2Cl-stab1}),
(\ref{eq:2Cl-stab2}), and (\ref{eq:2Cl-stab3}). The practical usefulness
of these conditions is illustrated later in Sec.~\ref{sec:Numerics}
by determining the region in the parameter space with the coexisting
stable two-cluster states. Recall, that the two-cluster solution (\ref{eq:2Cl-state})
corresponds to a fixed point $\delta_{\ast}=Y_{N_{1}}\left(\delta_{\ast}\right).$ 

Let us investigate the effect of small perturbations to (\ref{eq:2Cl-state}).
Because of the original system's symmetry with respect to index permutations,
it suffices to consider perturbations which do not change the phase-order
(see Appendix \ref{sub:return-map} and \citep{Goel2002}). Here,
these admissible perturbations are of the form\begin{align*}
 & \eta=\left(\eta_{1},\dots,\eta_{N_{1}},\eta_{N_{1}+1},\dots,\eta_{N}\right),\\
 & \eta_{1}\ge\dots\ge\eta_{N_{1}}\ge0,\ \eta_{N_{1}+1}\ge\dots\ge\eta_{N}=0.\end{align*}
To avoid tedious calculations, we do not perform a full linearization
of the return map and calculation of all multipliers. Instead, we
will provide estimates, that are sharp for $N\to\infty,$ for the
largest multipliers in the following invariant subspaces of admissible
perturbations:\begin{align*}
\mathcal{V}_{1} & =\left\{ \left(\eta_{1},\dots,\eta_{N_{1}-1},0,\dots,0\right)\,\mid\,\eta_{1}\ge\dots\ge\eta_{N_{1}}=0\right\} ,\\
\mathcal{V}_{2} & =\left\{ \left(0,\dots,0,\eta_{N_{1}+1},\dots,\eta_{N-1},0\right)\,\right|\\
 & \qquad\qquad\qquad\qquad\qquad\left.\eta_{N_{1}+1}\ge\dots\ge\eta_{N}=0\right\} ,\\
\mathcal{V}_{3} & =\left\{ \left(\eta_{1},\dots,\eta_{N_{1}},0,\dots,0\right)\,\mid\,\eta_{1}=\dots=\eta_{N_{1}}=\delta\ge0\right\} .\end{align*}
Here, $\mathcal{V}_{1}$ and $\mathcal{V}_{2}$ contain \emph{intra}-cluster
perturbations, which only affect phases inside one cluster, and $\mathcal{V}_{3}$
contains the perturbation of the cluster distance, i. e. the \emph{inter}-cluster
perturbations. Since the direct product of these invariant subspaces
contains all admissible perturbations, the linear stability of the
full system combines from stability in the subspaces.

Let us first turn to the intra-cluster perturbations in $\mathcal{V}_{1}$
and $\mathcal{V}_{2}.$ We introduce the cluster-widths for the state
$\left(\varphi_{1},...,\varphi_{N}\right)$ as\[
W_{1}=\varphi_{1}-\varphi_{N_{1}}\mbox{ and }W_{2}=\varphi_{N_{1}+1}-\varphi_{N}.\]
In the perfect two-cluster state (\ref{eq:2Cl-state}) we have $W_{1}=W_{2}=0.$
If this state is perturbed along $\mathcal{V}_{1},$ the cluster-width
of the first cluster becomes positive, i. e. $W_{1}=\eta_{1}\mbox{ and }W_{2}=0.$
Analogously, perturbations along $\mathcal{V}_{2}$ result in $W_{1}=0\mbox{ and }W_{2}=\eta_{N_{1}+1}.$ 

First, let us consider the perturbation within $\mathcal{V}_{1}$
when the first cluster with $N_{1}$ elements is perturbed. The intra-stability
of one cluster is determined by the change of its width during one
period, i.e. one application of the return map. These changes effectively
take place only at two events, i. e. at the crossing of the threshold
either by the first cluster itself or by the second cluster. Denote
the maximal possible change of the width of the perturbed cluster
during its own crossing as $\Delta_{1}^{N_{1}}\left(\varepsilon\right),$
where $\varepsilon\ge0$ is the width before the burst. This is, an
initial width $\varepsilon$ results in a new width $\varepsilon+\Delta_{1}^{N_{1}}\left(\varepsilon\right)$
after the burst. Since the other cluster is not involved in this process,
we can treat it as a burst of a single cluster in a system of $N_{1}$
oscillators. Sections \ref{sec:one-cluster} and \ref{sub:bifurcations-of-the-synchronous-state}
showed that, the maximal $\Delta_{1}^{N_{1}}\left(\varepsilon\right)$
is realized either by the perturbation $\varepsilon=\eta_{1}>\eta_{2}=\dots=\eta_{N_{1}}=0$
or by $\varepsilon=\eta_{1}=\dots=\eta_{N_{1}-1}>\eta_{N_{1}}=0,$
depending on the values of $Z^{\prime}\left(0\right)$ and $Z^{\prime}\left(2\pi\right).$

Similarly, we denote changes of the cluster's width at the crossing
of the other cluster as $\Delta_{2}^{N-N_{1}}\left(\varepsilon\right),$
where $\varepsilon\ge0$ is as before and $N-N_{1}$ is the number
of elements in the other, unperturbed cluster. Note that $\Delta_{2}^{N-N_{1}}\left(\varepsilon\right)$
is the same for all intra-cluster perturbations of magnitude $\varepsilon$,
since the phase order is preserved and solely the distance between
the first and the last phases $\varphi_{1}-\varphi_{N_{1}}$ in the
cluster matters.

The maximal width for perturbations of the first cluster after one
return is then\[
W_{1}\left(\varepsilon\right)=\varepsilon+\Delta_{1}^{N_{1}}\left(\varepsilon\right)+\Delta_{2}^{N-N_{1}}\left(\varepsilon+\Delta_{1}^{N_{1}}\left(\varepsilon\right)\right)\]
with $\varepsilon=\eta_{1}.$ 

Similarly, the width of the second cluster (within the invariant subspace
$\mathcal{V}_{2}$) changes as \[
W_{2}\left(\varepsilon\right)=\varepsilon+\Delta_{2}^{N_{1}}\left(\varepsilon\right)+\Delta_{1}^{N-N_{1}}\left(\varepsilon+\Delta_{2}^{N_{1}}\left(\varepsilon\right)\right),\]
for $\varepsilon=\eta_{N_{1}+1}.$ Stability conditions with respect
to all possible intra-cluster perturbations (within the subspace $\mathcal{V}_{1}\oplus\mathcal{V}_{2}$)
are then given by $W_{1}^{\prime}\left(0\right)<1$ and $W_{2}^{\prime}\left(0\right)<1,$
i. e.\begin{align}
W_{1}^{\prime}\left(0\right)=1+\left(\Delta_{1}^{N_{1}}\right)^{\prime}\left(0\right)+\left(\Delta_{2}^{N-N_{1}}\right)^{\prime}\left(0\right)\nonumber \\
+\left(\Delta_{2}^{N-N_{1}}\right)^{\prime}\left(0\right)\cdot\left(\Delta_{1}^{N_{1}}\right)^{\prime}\left(0\right) & <1,\label{eq:DW1-le-1}\\
W_{2}^{\prime}\left(0\right)=1+\left(\Delta_{2}^{N_{1}}\right)^{\prime}\left(0\right)+\left(\Delta_{1}^{N-N_{1}}\right)^{\prime}\left(0\right)\nonumber \\
+\left(\Delta_{1}^{N-N_{1}}\right)^{\prime}\left(0\right)\cdot\left(\Delta_{2}^{N_{1}}\right)^{\prime}\left(0\right) & <1.\label{eq:DW2-le-1}\end{align}
The stability analysis of the synchronous solution in sections \ref{sec:one-cluster}
and \ref{sub:bifurcations-of-the-synchronous-state}, applied to a
population of $n$ ($n=N_{1}$ or $n=N-N_{1}$) oscillators, implies
that $\Delta_{1}^{n}\left(\varepsilon\right)$ satisfies\begin{align*}
1+\left(\Delta_{1}^{n}\right)^{\prime}\left(0\right)= & \left(1+\frac{\varkappa}{N}\min\left(Z^{\prime}\left(0\right),Z^{\prime}\left(2\pi\right)\right)\right)^{1}\\
 & \times\left(1+\frac{\varkappa}{N}\max\left(Z^{\prime}\left(0\right),Z^{\prime}\left(2\pi\right)\right)\right)^{n-1},\end{align*}
see (\ref{eq:DYN1-in-0}). For large populations, this reads\begin{align}
1+\left(\Delta_{1}^{n}\right)^{\prime}\left(0\right) & \approx\exp\left(\varkappa p\max\left(Z^{\prime}\left(0\right),Z^{\prime}\left(2\pi\right)\right)\right),\label{eq:Delta1Nj-prime}\end{align}
where $p=n/N.$ This is obtained as in (\ref{eq:limitN-of-YN1}).
The change of the width of a perturbed cluster at position $\varphi=\delta,$
that is induced by the burst of another unperturbed cluster is given
as (see Appendix \ref{sub:Approx-of-repetitive-firing} for details)
\begin{align}
\Delta_{2}^{n}\left(\varepsilon\right) & =\sum_{k=0}^{n-1}\frac{\varkappa}{N}\left(Z\left(\mu^{k}\left(\delta+\varepsilon\right)\right)-Z\left(\mu^{k}\left(\delta\right)\right)\right).\label{eq:DeltaNj2-sum}\end{align}
 Hence,\begin{align}
\left(\Delta_{2}^{n}\right)^{\prime}\left(0\right) & =\sum_{k=0}^{n}\frac{\varkappa}{N}Z^{\prime}\left(\mu^{k}\left(\delta\right)\right)\left(\mu^{k}\right)^{\prime}\left(\delta\right).\label{eq:DeltaNj2-prime}\end{align}
One can approximate (see Appendix \ref{sub:Approx-of-repetitive-firing})
this sum in the limit $N\to\infty$ by a simpler expression using
the solution $\vartheta\left(r,\varphi\right)$ to the initial value
problem\begin{equation}
\begin{array}{rcl}
\frac{\partial\vartheta}{\partial r}\left(r,\varphi\right) & = & \varkappa Z\left(\vartheta\left(r,\varphi\right)\right),\\
\vartheta\left(0,\varphi\right) & = & \varphi.\end{array}\label{eq:IVP-vartheta}\end{equation}
As argued in the Appendix \ref{sub:Approx-of-repetitive-firing},
the function $\vartheta\left(r,\varphi\right)$ is a smooth approximation
of $\mu^{r\cdot N}\left(\varphi\right).$ Using this, we get\begin{equation}
\left(\Delta_{2}^{n}\right)^{\prime}\left(0\right)\approx\frac{Z\left(\vartheta\left(p,\delta\right)\right)-Z\left(\delta\right)}{Z\left(\delta\right)},\label{eq:DeltaNj2-prime-approx}\end{equation}
 where $p=n/N$ and $\delta$ is the position of the perturbed cluster,
when the unperturbed cluster crosses the threshold. 

Using (\ref{eq:DW1-le-1}), (\ref{eq:Delta1Nj-prime}) and (\ref{eq:DeltaNj2-prime-approx}),
we obtain for the two-cluster state (\ref{eq:2Cl-state}) of large
populations \begin{align*}
W_{1}^{\prime}\left(0\right)\approx & \exp\left(\varkappa p\max\left(Z^{\prime}\left(0\right),Z^{\prime}\left(2\pi\right)\right)\right)\\
 & \times\left(1+\frac{Z\left(\delta_{\ast}\right)-Z\left(2\pi-\vartheta\left(p,2\pi-\delta_{\ast}\right)\right)}{Z\left(2\pi-\vartheta\left(p,2\pi-\delta_{\ast}\right)\right)}\right),\end{align*}
where $p=N_{1}/N.$ That is, the condition for the intra-stability
of the first cluster, (\ref{eq:DW1-le-1}), leads to:\begin{align}
 & \exp\left(\varkappa p\max\left(Z^{\prime}\left(0\right),Z^{\prime}\left(2\pi\right)\right)\right)\nonumber \\
 & \times\left(1+\frac{Z\left(\delta_{\ast}\right)-Z\left(2\pi-\vartheta\left(p,2\pi-\delta_{\ast}\right)\right)}{Z\left(2\pi-\vartheta\left(p,2\pi-\delta_{\ast}\right)\right)}\right)<1,\label{eq:2Cl-stab1}\end{align}
 and for the second cluster, (\ref{eq:DW2-le-1}) reads \begin{align}
 & \exp\left(\varkappa\left(1-p\right)\max\left(Z^{\prime}\left(0\right),Z^{\prime}\left(2\pi\right)\right)\right)\nonumber \\
 & \times\left(1+\frac{Z\left(\vartheta\left(p,2\pi-\delta_{\ast}\right)\right)-Z\left(2\pi-\delta_{\ast}\right)}{Z\left(2\pi-\delta_{\ast}\right)}\right)<1.\label{eq:2Cl-stab2}\end{align}
 The stability of perturbations in $\mathcal{V}_{3},$ i. e. of the
cluster distance, corresponds to the stability of the cluster position
$\delta_{\ast}$ as a fixed point of $Y_{N_{1}},$ i. e. $\left|Y_{N_{1}}^{\prime}\left(\delta_{\ast}\right)\right|<1.$
For $N\to\infty,$ this condition can be rewritten as\begin{align}
\frac{Z\left(\delta_{\ast}\right)\cdot Z\left(\vartheta\left(p,2\pi-\delta_{\ast}\right)\right)}{Z\left(2\pi-\vartheta\left(p,2\pi-\delta_{\ast}\right)\right)\cdot Z\left(2\pi-\delta_{\ast}\right)} & <1.\label{eq:2Cl-stab3}\end{align}
For more details see Appendix \ref{sub:Approx-of-repetitive-firing}. 

Note that, for $\varkappa\to0,$ i. e. very weak coupling, we can
simplify (\ref{eq:2Cl-stab1})--(\ref{eq:2Cl-stab3}) to\begin{align}
\left(1-p\right)Z^{\prime}\left(\delta_{0}\right)+p\max\left(Z^{\prime}\left(0\right),Z^{\prime}\left(2\pi\right)\right) & <0,\label{eq:2Cl-stab-weak-coupling1}\\
pZ^{\prime}\left(2\pi-\delta_{0}\right)+\left(1-p\right)\max\left(Z^{\prime}\left(0\right),Z^{\prime}\left(2\pi\right)\right) & <0,\label{eq:2Cl-stab-weak-coupling2}\\
p\cdot Z^{\prime}\left(2\pi-\delta_{0}\right)+\left(1-p\right)Z^{\prime}\left(\delta_{0}\right) & <0,\label{eq:2Cl-stab-weak-coupling3}\end{align}
where $\delta_{0}=\lim_{\varkappa\to0}\delta_{\ast}$ -- see Appendix
\ref{sub:Weak-coupling}. These latter conditions match exactly the
conditions, that were obtained by Hansel et. al. in \citep{Hansel1993}
for the linear stability in the averaged model \[
\dot{\varphi}_{i}\left(t\right)=1+\frac{\varkappa}{N}\sum_{j=1}^{N}Z\left(\varphi_{i}\left(t\right)-\varphi_{j}\left(t\right)\right).\]
As in the averaged model, in the pulse-coupled case it is possible
to encounter multistability in between stable cluster-states and the
stable synchronous solution.

\subsection{A special case}

Let us now consider the situation\begin{equation}
Z^{\prime}\left(0\right)=Z^{\prime}\left(2\pi\right)=0.\label{eq:assumption-Zprime0}\end{equation}
This might seem degenerate, but in neuron models as well as in experimental
neurophysiology, this property is not unusual -- review the examples
in Fig.~\ref{fig:PRC-examples} and note that the derivatives are
indeed zero at $\varphi=0$ and $\varphi=2\pi$. We picked a simple
example of this kind to illustrate our results numerically in the
Sec.~\ref{sec:Numerics}. Given (\ref{eq:assumption-Zprime0}), formula
(\ref{eq:linstab1cl}) does not supply any information about the stability
of the synchronous solution, because one finds that for arbitrary
$N_{1},$

\[
\left(1+\frac{\varkappa}{N}Z^{\prime}\left(0\right)\right)^{N_{1}}\left(1+\frac{\varkappa}{N}Z^{\prime}\left(2\pi\right)\right)^{N-N_{1}}=1\]
and all multipliers are degenerate and equal to 1. Since the linear
stability analysis does not provide any results, one must consider
terms of higher order to complete a stability analysis for this case.
We will go one step in this direction by investigating the nonlinear
stability of the synchronous solution along the invariant split states
as (\ref{eq:two-cluster}) and point out that similar mechanisms may
participate in a destabilization as in the previous section. Consider
quadratic terms along the split directions. One finds:\begin{align}
Y_{N_{1}}^{\prime\prime}\left(0\right) & =\left(N-N_{1}\right)\frac{\varkappa}{N}Z^{\prime\prime}\left(0\right)-N_{1}\frac{\varkappa}{N}Z^{\prime\prime}\left(2\pi\right),\label{eq:D2-YN1-0}\\
Y_{N_{1}}^{\prime\prime}\left(2\pi\right) & =\left(N-N_{1}\right)\frac{\varkappa}{N}Z^{\prime\prime}\left(2\pi\right)-N_{1}\frac{\varkappa}{N}Z^{\prime\prime}\left(0\right)\nonumber \\
 & =-Y_{N-N_{1}}^{\prime\prime}\left(0\right).\label{eq:D2-YN1-2pi}\end{align}
Up to the leading terms, the dynamics governing small distances between
two-clusters is given now by 

\[
\delta\to\delta+\frac{1}{2}Y''_{N_{1}}\left(0\right)\delta^{2},\]
instead of (\ref{eq:linz}). Thus, we have an analogous situations
as in Sec.~\ref{sub:bifurcations-of-the-synchronous-state}, Fig.~\ref{fig:fixpt-and-cobweb-for-Y}.
The synchronous state is stable under conditions \[
Y''_{N_{1}}\left(0\right)<0,\quad Y''_{N_{1}}(2\pi)>0.\]

Bifurcations from stable synchronous state to two-cluster-states occur
in the same way as before. The conditions for the stability of two-cluster
states (\ref{eq:2Cl-stab1})--(\ref{eq:2Cl-stab2}) from Sec.~\ref{sub:2Cl-stab-formula}
reduce to \begin{align}
\frac{Z\left(\delta_{\ast}\right)}{Z\left(2\pi-\vartheta\left(p,2\pi-\delta_{\ast}\right)\right)}<1 & \mbox{ and }\frac{Z\left(\vartheta\left(p,2\pi-\delta_{\ast}\right)\right)}{Z\left(2\pi-\delta_{\ast}\right)}<1,\label{eq:2Cl-stab-for-DZ0-0}\end{align}
 and imply (\ref{eq:2Cl-stab3}) in the present case. Once more, for
weak coupling, these conditions takes a simpler form, that is \begin{align}
Z^{\prime}\left(\delta_{0}\right)<0 & \mbox{ and }Z^{\prime}\left(2\pi-\delta_{0}\right)<0,\label{eq:2Cl-stab-for-DZ0-weak-coupl}\end{align}
where $\delta_{0}=\lim_{\varkappa\to0}\delta_{\ast}.$

\section{Numerics\label{sec:Numerics}}

In this section we provide a simple example, where branches of initially
unstable two-clusters emerge and eventually stabilize for increasing
$\beta,$ as predicted by our analytics (see Fig.~\ref{fig:stabdots}).
Moreover, we will illustrate the stable homoclinic structure leading
to intermittent synchronization (see Fig.~\ref{fig:1Cl-homoclinic-and-spread}).

In order to detect the appearance of one- or two-cluster states, we
compute the order parameters \begin{equation}
R_{1}(t)=\left|\frac{1}{N}\sum_{j=1}^{N}e^{i\varphi_{j}(t)}\right|\label{op1}\end{equation}
 and \begin{equation}
R_{2}(t)=\left|\frac{1}{N}\sum_{j=1}^{N}e^{i2\varphi_{j}(t)}\right|.\label{op2}\end{equation}
A perfect one-cluster state is characterized by $R_{1}=R_{2}=1$ and
a perfect antiphase two-cluster is characterized by $R_{1}=0$ and
$R_{2}=1$. We present results of simulations for $\varkappa=0.5,$
but qualitatively we observe similar behavior for a broad range of
$\varkappa>0.$

\begin{figure}
\includegraphics[width=0.45\textwidth]{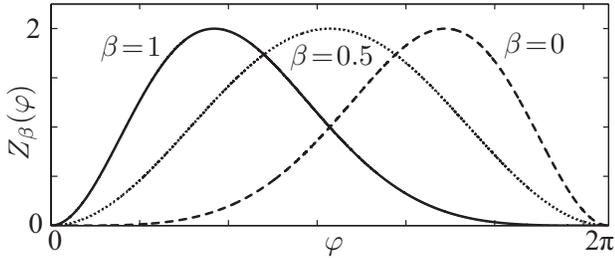}\caption{\label{fig:Zbeta-family}Family of unimodal PRCs $Z_{\beta}(\varphi)$
that were used in simulations, see (\ref{eq:Zbeta-family}).}

\end{figure}

Consider the following family of PRCs (see Fig.~\ref{fig:Zbeta-family}):\begin{align}
Z_{\beta}\left(\varphi\right) & =1-\cos\left(\chi_{\beta}\left(\varphi\right)\right)\mbox{, for }\beta\in\left[0,1\right],\label{eq:Zbeta-family}\\
\chi_{\beta}\left(\varphi\right) & =\frac{\left(1-2\beta\right)}{2\pi}\varphi^{2}+2\beta\varphi.\nonumber \end{align}

\begin{figure}
\includegraphics[width=0.45\textwidth]{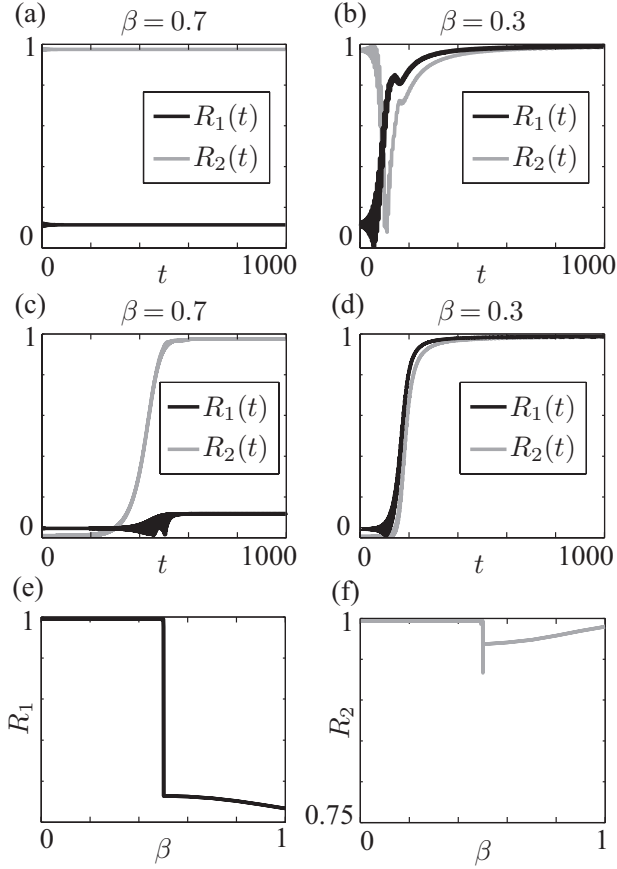}\caption{\label{OPtimelines}Asymptotic behavior of the order parameters. Charts
(a) and (b) show $R_{1}(t)$ and $R_{2}(t)$ for a trajectory starting
in a vicinity of the two-cluster state. The middle panel (c) and (d)
belong to a trajectories starting in a vicinity of the splay state.
Left charts (a) and (c) correspond to the parameter value $\beta=0.7$,
where the two-cluster state is attracting and (b) and (d) to $\beta=0.3$,
where one-cluster state is attracting. \label{OPvsbeta} In (e) and
(f): dependence of the asymptotic values for the order parameter $R_{1}$
and $R_{2}$ on $\beta$. For the most values of $\beta$, except
$\beta=0.5$, the order parameters tend to some constant value, when
initialized near the splay state or the symmetric two-cluster state. }

\end{figure}

As shown in Fig.~\ref{OPtimelines}, we observe two qualitatively
different types of behavior depending on parameter $\beta$. For $\beta<0.5$,
i.e. when the maximum of the PRC is shifted to the right (see Fig.~\ref{fig:Zbeta-family}),
the one-cluster state acts attracting on most initial states; for
$\beta>0.5$ the maximum of the PRC is shifted to the left and two-cluster
states become attracting. We have chosen initial conditions in a vicinity
of a two-cluster state in Fig.~\ref{OPtimelines}(a) and (b), therefore
the initial values of the order parameters are $R_{1}\approx0$ and
$R_{2}\approx1$. Figure \ref{OPtimelines}(b) shows how the instability
of the two-cluster state implies desynchronization transient, after
which the system is attracted to a synchronous one-cluster state.
Similar behavior occurs for other initial conditions. Figure~\ref{OPtimelines}(c)
and (d) illustrate the order parameters behavior for initial conditions
close to the splay state (a state, where the phases are distributed).
The initial values for the order parameters in the splay state are
close to zero, but after a transient, they approach again the same
asymptotic values as in (a) and (b).

A more complicated behavior occurs for the intermediate value of the
parameter $\beta=0.5$, i.e. when the PRC is symmetric. In this case,
the order parameters $R_{1}(t)$ and $R_{2}(t)$ do not approach some
asymptotic constant values but remain periodic in time. As a result,
the maximum asymptotic values of both $R_{1}$ and $R_{2}$ do not
coincide with the corresponding minimum values. This type of behavior
is observed for a very small parameter interval of magnitude $O\left(\frac{1}{N}\right)$
around $\beta\approx0.5$. It is worth mentioning, that $Z_{0.5}\left(\varphi\right)=1-\cos\left(\varphi\right)$
was proposed as PRC for oscillators near a saddle-node bifurcation
on a periodic orbit \citep{Ermentrout1996,Brown2004}. Since $\beta=0.5$
is a critical value for our model, we advise caution for treating
this case as exemplary in investigations. Figure~\ref{OPvsbeta}
(e) and (f) summarize the behavior of the order parameters for different
$\beta$.

\subsection{Cluster-stability\label{sub:Numerical-cluster-stab}}

\begin{figure}
\includegraphics[width=0.45\textwidth]{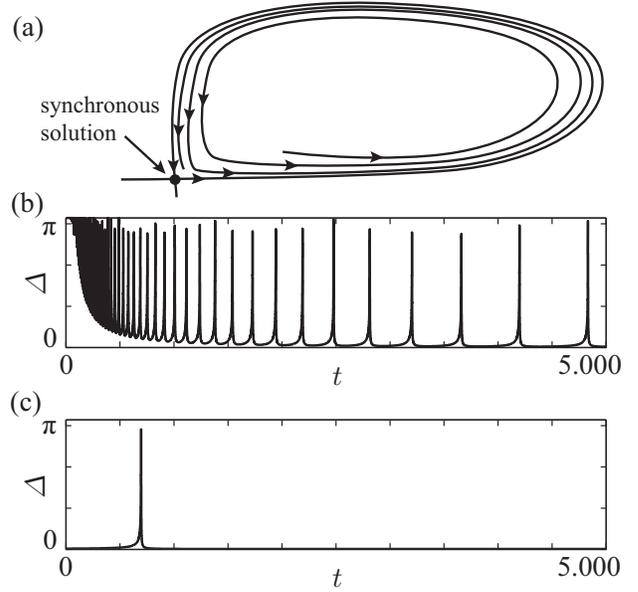}\caption{\label{fig:1Cl-homoclinic-and-spread}Homoclinic connection of the
one-cluster. In (a): Synchronous solution (i.e. one-cluster state)
as a saddle point in the phase space with a homoclinic loop. In (b)
and (c): Width of the cluster $\Delta(t)=\max_{1\le i,j\le N}\left\{ \left|\varphi_{i}(t)-\varphi_{j}(t)\right|\right\} $
as a function of time. Chart (a) shows the behavior along the orbit
started at an initial condition close to the splay state (far from
the one-cluster) and (b) shows the behavior along the orbit started
close to a split-state with homoclinic connection as in Fig.~\ref{fig:fixpt-and-cobweb-for-Y}
(b). The growing intervals between peaks in (a) indicate the existence
of a stable homoclinic orbit. }

\end{figure}

Since $Z_{\beta}^{\prime}\left(0\right)=Z_{\beta}^{\prime}\left(2\pi\right)=0,$
for all $\beta\in\left[0,1\right],$ we have \[
Y_{\beta,N_{1}}^{\prime}\left(0\right)=Y_{\beta,N_{1}}^{\prime}\left(2\pi\right)=1\mbox{, for all }N_{1}=1,...,N-1.\]
 This means, expressions (\ref{eq:D2-YN1-0}) and (\ref{eq:D2-YN1-2pi})
become relevant to stability of the synchronous solutions. In our
example, they read\begin{align*}
Y_{\beta,N_{1}}^{\prime\prime}\left(0\right) & =4\varkappa\left(\left(1-p\right)\beta^{2}-p\left(1-\beta\right)^{2}\right),\\
Y_{\beta,N_{1}}^{\prime\prime}\left(2\pi\right) & =4\varkappa\left(\left(1-p\right)\left(1-\beta\right)^{2}-p\beta^{2}\right),\end{align*}
 where $p=N_{1}/N.$ For large populations, the synchronous solution
is unstable for any $\beta.$ Nevertheless, numerics show an attraction
to the synchronous state, which can be explained by the existence
of a stable homoclinic connection of the one-cluster (\citep{Luecken2011},
see Fig.~\ref{fig:1Cl-homoclinic-and-spread}). %

\begin{figure}
\includegraphics[width=0.45\textwidth]{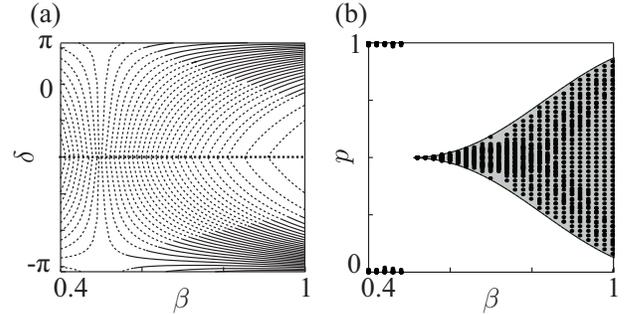}\caption{\label{fig:stabdots}Stability and existence of two-cluster states.
(a) A cascade of pitchfork bifurcations gives birth to fixed points
of $Y_{N_{1}}$ and $Y_{N-N_{1}}.$ Solid lines denote stable two-cluster
stationary states and dashed - unstable. The lines are shown only
for selected values of $p=N_{1}/N.$ Markers in Fig.~(b) shows which
two-clusters are stable in dependence on $\beta$ and $p$ (observed
numerically). The value $p=0.5$ corresponds to the symmetric cluster
and $p\ne0.5$ to nonsymmetric clusters. The shaded region in (b)
indicates the analytic prediction for the stability of the two-cluster
states by (\ref{eq:2Cl-stab-for-DZ0-0}).}

\end{figure}

At $\beta\approx0.5,$ the symmetric two-cluster state gains stability
and subsequently, with increasing $\beta,$ asymmetric cluster-states
emerge when $Y_{\beta,N_{1}}^{\prime\prime}\left(0\right)$ changes
sign and gain stability when (\ref{eq:2Cl-stab-for-DZ0-0}) is fulfilled.
The analytic predictions from Sec.~\ref{sub:2Cl-stab-formula} match
the numerical results. More specifically, in Fig.~\ref{fig:stabdots}(a)
one observes pitchfork bifurcations of the synchronous state ($\delta=0$)
leading to the appearance of two-cluster states. These two-cluster
states are initially unstable but, with increasing $\beta$, they
gain stability creating a large set of coexisting stable two-clusters.
Figure \ref{fig:stabdots}(b) shows the stable clusters versus $\beta$.
Solid line corresponds to the analytically predicted stability domain
(\ref{eq:2Cl-stab-for-DZ0-0}) and the dots are observed clusters
from numerical calculations. %

\section{Discussion }

In this paper, we have considered the following question: How the
individual properties of oscillators in a globally coupled system
influence the formations of one- and two-cluster states? Working in
the framework of pulse-coupled oscillators, the adjustable parameter
for our study was chosen to be the PRC. 

Starting from the results of Goel and Ermentrout \citep{Goel2002}
about the stability of one-cluster state, we have extended these results
to the case when the first derivatives of the PRC at the threshold
equal to zero. We have derived a one-dimensional map that describes
the dynamics in invariant subspaces corresponding to various two-cluster
states. By means of this map we identified pitchfork bifurcations
of the one-cluster state, which lead to the emergence of periodic
two-cluster states. For large populations, we provide explicit conditions
for the linear stability of these states. The analysis of the one-dimensional
map also reveals that a homoclinic connection to the synchronous one-cluster
state is generic in the considered class of systems. Moreover, numerical
analysis shows that a set of homoclinic connections can be globally
attracting.

For the numerical illustration, we have chosen systems with positive,
unimodal PRCs, which turn to zero at the threshold together with its
first derivatives. This corresponds to the excitatory coupling. By
varying the shape of this PRCs, we observe how the initially globally
attracting one-cluster undergoes a sequence of bifurcations in which
two-cluster states emerge and later on stabilize. This leads to the
coexistence of multiple stable two-cluster states. The latter phenomenon
resembles qualitatively the Eckhaus phenomenon \citep{Eckhaus1965,Wolfrum2006}.

\section*{Acknowledgments}

This work was supported by Deutsche Forschungsgemeinschaft in the
framework of SFB910 under the project A3 (L.L.) and Research Center
Matheon under the project D21 (S.Y.).

\section{Appendix }

\subsection{Formal definition of the return map\label{sub:return-map}}

In this section, we derive the discrete return map for the system
(\ref{eq:system-def-1}). Let us introduce the state-spaces \[
X_{j}=\left\{ \left(\varphi_{1},...,\varphi_{N}\right)\in\left[0,2\pi\right]\mid\varphi_{\left[1+j\right]}\ge...\ge\varphi_{\left[N+j\right]}=0\right\} ,\]
where $\left[n\right]=n\,\mbox{mod}\, N.$ In particular\[
X_{N}=X_{0}=\left\{ \left(\varphi_{1},...,\varphi_{N}\right)\in\left[0,2\pi\right]\mid\varphi_{1}\ge...\ge\varphi_{N}=0\right\} .\]
Since all oscillators are taken to be identical and homogeneously
coupled, the system is symmetric with respect to permutations of indexes
($\mathfrak{S}_{N}$-symmetry). Therefore, without loss of generality,
let us consider initial states in $X_{0}.$ The return map $K:\, X_{0}\mapsto X_{0}$
is composed of $N$ maps $k_{j}:X_{j-1}\mapsto X_{j}$, $j=1,...,N,$
defined as\begin{align*}
 & k_{j}\left(\varphi_{1},...,\varphi_{j},...,\varphi_{N}\right):=\\
 & \left(\varphi_{1}+2\pi-\varphi_{j}+\frac{\varkappa}{N}Z\left(\varphi_{1}+2\pi-\varphi_{j}\right),\dots\right.\\
 & \left.\dots,\varphi_{N}+2\pi-\varphi_{j}+\frac{\varkappa}{N}Z\left(\varphi_{N}+2\pi-\varphi_{j}\right)\right).\end{align*}
where the $j-th$ The map $k_{j}$ describes the firing of the $j$-th
oscillator, assuming it has maximal phase. Under the assumption that
the phase order is preserved, i.e. (\ref{eq:phaseordercond}) holds,
the return map is given by\[
K=k_{N}\circ...\circ k_{1}:\, X_{0}\to X_{1}\to...\to X_{N}=X_{0}.\]
Note that each $k_{j}$ is smooth. Linearizations of $K$ capture
perturbations, which keep the presumed phase-ordering. Because of
the system's symmetry, it suffices to consider these for a determination
of linear stability (see also \citep{Goel2002}).

\subsection{Approximation of repetitive firing\label{sub:Approx-of-repetitive-firing}}

Assume that (\ref{eq:phaseordercond}) holds. Then, the phase order
is preserved. Consider a perturbed cluster, $\varphi_{1}\ge\dots\ge\varphi_{N_{1}},$
with $N_{1}=p\cdot N$ and \[
\varphi_{1}=\delta+\varepsilon,\quad\varphi_{N_{1}}=\delta.\]
The change of the width of this cluster during the threshold crossing
of another unperturbed cluster \[
\varphi_{N_{1}+1}=\dots=\varphi_{N}=2\pi\]
with $N_{2}=\left(1-p\right)N$ oscillators is given by the change
of the distance between the leading oscillator $\varphi_{1}$ and
the last oscillator $\varphi_{N_{1}}.$ The new position $\varphi_{j}^{+}$
of a single oscillator $\varphi_{j}$ of the perturbed cluster is
given as \begin{align*}
\varphi_{j}^{+} & =\mu^{N_{2}}(\varphi_{j})=\underbrace{\mu(\mu(\mu(\cdots\mu}_{N_{2}}(\varphi_{j})\cdots)))\\
 & =\varphi_{j}+\sum_{k=0}^{N_{2}-1}\frac{\varkappa}{N}Z\left(\mu^{k}\left(\varphi_{j}\right)\right).\end{align*}
Thus, the change of width of the perturbed cluster is\begin{align*}
\varphi_{1}^{+}-\varphi_{N_{1}}^{+}= & \left(\varepsilon+\delta+\sum_{k=0}^{N_{2}-1}\frac{\varkappa}{N}Z\left(\mu^{k}\left(\delta+\varepsilon\right)\right)\right)\\
 & -\left(\delta+\sum_{k=0}^{N_{2}-1}\frac{\varkappa}{N}Z\left(\mu^{k}\left(\delta\right)\right)\right)\\
= & \varepsilon+\sum_{k=0}^{N_{2}-1}\frac{\varkappa}{N}\left(Z\left(\mu^{k}\left(\delta+\varepsilon\right)\right)-Z\left(\mu^{k}\left(\delta\right)\right)\right).\end{align*}
In this way, we arrive at (\ref{eq:DeltaNj2-sum}).

Let us introduce a function, $\vartheta:\left[0,1\right]\times\left[0,2\pi\right]\to\left[0,2\pi\right],$
that approximates the iterations of $\mu$ as follows: \[
\vartheta\left(r,\varphi\right)\approx\mu^{k}\left(\varphi\right)\mbox{, for }r=\frac{k}{N}.\]
 Then,\begin{align*}
\vartheta\left(r+\frac{1}{N},\varphi\right)-\vartheta\left(r,\varphi\right) & \approx\mu^{r\cdot N+1}\left(\varphi\right)-\mu^{rN}\left(\varphi\right)\\
 & =\frac{\varkappa}{N}Z\left(\mu^{rN}\left(\varphi\right)\right)\\
 & \approx\frac{\varkappa}{N}Z\left(\vartheta\left(r,\varphi\right)\right),\end{align*}
or equivalently,

\[
\frac{\vartheta\left(r+\frac{1}{N},\varphi\right)-\vartheta\left(r,\varphi\right)}{1/N}\approx\varkappa Z\left(\vartheta\left(r,\varphi\right)\right).\]
Therefore, for large $N$, $\vartheta\left(r,\varphi\right)$ is a
solution to the initial value problem (\ref{eq:IVP-vartheta}). Further,
$\frac{\partial\vartheta}{\partial\varphi}\left(r,\varphi\right)$
solves the linear system \begin{align*}
\frac{\partial}{\partial r}\frac{\partial\vartheta}{\partial\varphi}\left(r,\varphi\right) & =\varkappa Z^{\prime}\left(\vartheta\left(r,\varphi\right)\right)\frac{\partial\vartheta}{\partial\varphi}\left(r,\varphi\right),\\
\frac{\partial\vartheta}{\partial\varphi}\left(0,\varphi\right) & =1.\end{align*}
Therefore,\begin{align*}
\frac{\partial\vartheta}{\partial\varphi}\left(r,\varphi\right) & =\exp\left(\int_{0}^{r}\varkappa Z^{\prime}\left(\vartheta\left(s,\varphi\right)\right)ds\right).\end{align*}
Taking into account the property (\ref{eq:IVP-vartheta}), we obtain
\begin{align}
\frac{\partial\vartheta}{\partial\varphi}\left(r,\varphi\right) & =\exp\left(\int_{\varphi}^{\vartheta(r,\varphi)}\frac{Z^{\prime}\left(\theta\right)}{Z\left(\theta\right)}d\theta\right)=\frac{Z\left(\vartheta\left(r,\varphi\right)\right)}{Z\left(\varphi\right)}.\label{eq:dvartheta-dphi}\end{align}
Now, we can simplify the expression (\ref{eq:DeltaNj2-prime}) by
means of $\vartheta$. For this, we substitute $\mu^{k}\left(\varphi\right)$
by $\vartheta\left(\frac{k}{N},\varphi\right)$ and the sum over $k$
by an integral over $r$ in (\ref{eq:DeltaNj2-prime}). As a result,
we obtain

\begin{align*}
 & \ \sum_{k=0}^{N_{2}-1}\frac{\varkappa}{N}Z^{\prime}\left(\mu^{k}\left(\delta\right)\right)\left(\mu^{k}\right)^{\prime}\left(\delta\right)\\
\approx & \ \int_{0}^{1-p}\varkappa Z^{\prime}\left(\vartheta\left(r,\delta\right)\right)\frac{d\vartheta}{d\varphi}\left(r,\delta\right)dr\\
= & \ \frac{Z\left(\vartheta\left(1-p,\delta\right)\right)-Z\left(\delta\right)}{Z\left(\delta\right)}.\end{align*}
 Thus, for large populations, (\ref{eq:DeltaNj2-prime}) leads to
(\ref{eq:DeltaNj2-prime-approx}). 

The inter-cluster-stability is given by the stability of $\delta_{\ast}$
as a fixed point of $Y_{p\cdot N}$, that is, $\left|Y_{p\cdot N}^{\prime}\left(\delta_{\ast}\right)\right|<1.$
In terms of $\vartheta\left(r,\varphi\right),$ we have $Y_{p\cdot N}\left(\delta\right)\approx\vartheta\left(1-p,2\pi-\vartheta\left(p,2\pi-\delta\right)\right)$
and fixed points $\delta_{\ast}=Y_{p\cdot N}\left(\delta_{\ast}\right)$
satisfy $\delta_{\ast}\approx\vartheta\left(1-p,2\pi-\vartheta\left(p,2\pi-\delta_{\ast}\right)\right).$
Thereby one obtains (\ref{eq:2Cl-stab3}).

\subsection{Weak coupling in large populations\label{sub:Weak-coupling}}

Assume that for sufficiently small $\varkappa,$ there exist fixed
points $\delta\left(\varkappa\right)=Y_{p\cdot N}\left(\delta\left(\varkappa\right),\varkappa\right),$
with $p\cdot N\in\left\{ 1,\dots,N\right\} .$ Further, assume that
the limit $\delta_{0}=\lim_{\varkappa\to0}\delta\left(\varkappa\right)$
exists. Actually, it is a generic property for any point $\delta_{0}\in\left(0,2\pi\right),$
that fulfills $\left(1-p\right)Z\left(\delta_{0}\right)=pZ\left(2\pi-\delta_{0}\right),$
that there exists a family of fixed points of $Y_{p\cdot N}$ with
$\delta_{0}=\lim_{\varkappa\to0}\delta\left(\varkappa\right).$

For small $\varkappa$, the linearization of (\ref{eq:2Cl-stab1})--(\ref{eq:2Cl-stab3})
in $\varkappa=0$ gives the stability conditions (\ref{eq:2Cl-stab-weak-coupling1})--(\ref{eq:2Cl-stab-weak-coupling3}).
For example, the linearization of (\ref{eq:2Cl-stab1}) is\begin{align}
0> & \frac{\partial}{\partial\varkappa}\Biggl(\exp\left(\varkappa p\max\left(Z^{\prime}\left(0\right),Z^{\prime}\left(2\pi\right)\right)\right)\nonumber \\
 & \times\left(1+\frac{Z\left(\delta_{\ast}\right)-Z\left(2\pi-\vartheta\left(p,2\pi-\delta_{\ast}\right)\right)}{Z\left(2\pi-\vartheta\left(p,2\pi-\delta_{\ast}\right)\right)}\right)\Biggr)\label{eq:weak-coupling-calculation-01}\end{align}
In the following, we add $\varkappa$ explicitly as an argument where
needed. Note that the two-cluster fixed points $\delta_{\ast}=\delta\left(\varkappa\right)$
are $\varkappa$-dependent as well as the function $\vartheta\left(r,\varphi\right)=\vartheta\left(r,\varphi,\varkappa\right)$
from the previous section and $Y_{p\cdot N}\left(\varphi\right)=Y_{p\cdot N}\left(\varphi,\varkappa\right).$
Using (\ref{eq:IVP-vartheta}), we obtain $\frac{\partial\vartheta}{\partial\varkappa}\left(0,\varphi,\varkappa\right)$
as a solution to \begin{align*}
\frac{\partial}{\partial r}\frac{\partial\vartheta}{\partial\varkappa}\left(r,\varphi,\varkappa\right)= & \ Z\left(\vartheta\left(r,\varphi,\varkappa\right)\right)\\
 & \ +\varkappa Z^{\prime}\left(\vartheta\left(r,\varphi,\varkappa\right)\right)\frac{\partial\vartheta}{\partial\varkappa}\left(r,\varphi,\varkappa\right),\\
\frac{\partial\vartheta}{\partial\varkappa}\left(0,\varphi,\varkappa\right)= & \ 0.\end{align*}
This can be solved explicitly as \begin{align}
\frac{d\vartheta}{d\varkappa}\left(r,\varphi,\varkappa\right) & =r\cdot Z\left(\vartheta\left(r,\varphi,\varkappa\right)\right).\label{eq:dtheta-dkappa}\end{align}
Using $\vartheta\left(r,\varphi,0\right)=\varphi$, (\ref{eq:dvartheta-dphi})
and (\ref{eq:dtheta-dkappa}), the condition (\ref{eq:weak-coupling-calculation-01})
in $\varkappa=0$ becomes\begin{align}
0>\  & p\max\left(Z^{\prime}\left(0\right),Z^{\prime}\left(2\pi\right)\right)\nonumber \\
+ & \frac{\partial}{\partial\varkappa}\left(\frac{Z\left(\delta\left(\varkappa\right)\right)-Z\left(2\pi-\vartheta\left(p,2\pi-\delta\left(\varkappa\right),\varkappa\right)\right)}{Z\left(2\pi-\vartheta\left(p,2\pi-\delta\left(\varkappa\right),\varkappa\right)\right)}\right)\nonumber \\
=\  & p\max\left(Z^{\prime}\left(0\right),Z^{\prime}\left(2\pi\right)\right)+\frac{pZ\left(2\pi-\delta_{0}\right)Z^{\prime}\left(\delta_{0}\right)}{Z\left(\delta_{0}\right)}.\label{eq:weak-coupling-calculation-02}\end{align}
Differentiation of the fixed point equation $\delta\left(\varkappa\right)=Y_{p\cdot N}\left(\delta\left(\varkappa\right),\varkappa\right)\approx\vartheta\left(1-p,2\pi-\vartheta\left(p,2\pi-\delta\left(\varkappa\right),\varkappa\right),\varkappa\right)$
in $\varkappa=0$ yields\begin{align}
0 & \approx\left(1-p\right)Z\left(\delta_{0}\right)-pZ\left(2\pi-\delta_{0}\right).\label{eq:condition-on-delta0}\end{align}
Inserting this into (\ref{eq:weak-coupling-calculation-02}) gives
(\ref{eq:2Cl-stab-weak-coupling1}). Similarly, one obtains (\ref{eq:2Cl-stab-weak-coupling2})
from (\ref{eq:2Cl-stab2}) and (\ref{eq:2Cl-stab-weak-coupling3})
from (\ref{eq:2Cl-stab3}).

\bibliographystyle{model1-num-names}
\bibliography{Clusters}

\end{document}